\documentclass[aps,prx,floatfix,twocolumn,showpacs,10pt,longbibliography]{revtex4-2}
\usepackage{amsmath}
\usepackage{url}
\usepackage[ruled,vlined]{algorithm2e}
\usepackage{algorithmic}
\usepackage{graphicx}
\usepackage{dcolumn}
\usepackage{bm}
\usepackage{amssymb}
\usepackage{rotating}
\usepackage[abs]{overpic}
\usepackage{xcolor}
\usepackage{tabularx}
\usepackage{hyperref}
\usepackage[utf8]{inputenc}
\usepackage{braket}
\usepackage{soul}
\usepackage{qcircuit}
\usepackage[english]{babel}

\DeclareUnicodeCharacter{2009}{}

\newcommand{\tc}{\textcolor{black}}

\newcommand{\tctwo}{\textcolor{black}}

\newcommand{\tcthree}{\textcolor{black}}

\newcommand{\tcfour}{\textcolor{black}}

\begin{document}

\title{Quantum Simulation of Quantum Phase Transitions Using the Convex Geometry of Reduced Density Matrices}
\author{Samuel Warren, LeeAnn M. Sager-Smith, and David A. Mazziotti}

\email{damazz@uchicago.edu}

\affiliation{Department of Chemistry and The James Franck Institute, The University of Chicago, Chicago, IL 60637}%

\date{Submitted November 9, 2021\tc{; Revised February 4, 2022}\tctwo{; Revised April 22, 2022}\tcthree{; Revised June 15, 2022}}


\begin{abstract}
Transitions of many-particle quantum systems between distinct phases at absolute-zero temperature, known as quantum phase transitions, require an exacting treatment of particle correlations. In this work, we present a general quantum-computing approach to quantum phase transitions that exploits the geometric structure of reduced density matrices.  While typical approaches to quantum phase transitions examine discontinuities in the order parameters, the origin of phase transitions---their order parameters and symmetry breaking---can be understood geometrically in terms of the set of two-particle reduced density matrices (2-RDMs).  The convex set of 2-RDMs provides a comprehensive map of the quantum system including its distinct phases as well as the transitions connecting these phases.  \tcthree{Because 2-RDMs can potentially be computed on quantum computers at non-exponential cost, even when the quantum system is strongly correlated, they are ideally suited for a quantum-computing approach to quantum phase transitions.}  We compute the convex set of 2-RDMs for a Lipkin-Meshkov-Glick spin model on IBM superconducting-qubit quantum processors.  \tcthree{Even though computations are limited to few-particle models due to device noise, comparisons with a classically solvable 1000-particle model reveal that the finite-particle quantum solutions capture the key features of the phase transitions including the strong correlation and the symmetry breaking.}
\end{abstract}

\maketitle

\section{Introduction}
Phase transitions such as the melting of ice arise from thermal fluctuations~\cite{gibbs_graphical_1873}. Even at absolute zero in the absence of thermal fluctuations, transitions between phases, known as quantum phase transitions (QPTs), can occur from quantum fluctuations arising from the uncertainty relations~\cite{sachdev_quantum_2011, landau_intersection_1965, maslov_nonanalytic_2009}. Importantly, these quantum fluctuations and their associated phase transitions are significant for a range of temperatures beginning at absolute zero. An understanding of these transitions is critical to addressing outstanding problems in the study of magnetic insulators~\cite{bitko_quantum_1996,zhang_topology-driven_2013,maiani_topological_2021}, electron gases~\cite{perdew_generalized_1996, sondhi_continuous_1997,schreiber_electronelectron_2018,nataf_rashba_2019}, heavy-fermion compounds~\cite{von_lohneysen_non-fermi-liquid_1998,knafo_antiferromagnetic_2009,ronning_electronic_2017}, and high temperature superconductors~\cite{fisher_quantum_1990, dagotto_correlated_1994,maple_high-temperature_1998,orenstein_advances_2000,sachdev_quantum_2000,soltan-panahi_quantum_2012, holmvall_broken_2018}.  Experimental realizations of QPTs often involve laser traps of individual atoms and ions~\cite{fisher_boson_1989, greiner_quantum_2002, islam_onset_2011, zhang_observation_2017, cai_observation_2021, ebadi_quantum_2021}, or the synthesis of exotic materials~\cite{ demko_disorder_2012, xing_ising_2017, saito_quantum_2018} requiring significant investment in experimental setups and synthesis techniques.  The advent of cloud-accessible quantum computing devices~\cite{noauthor_ibm_2021}, which allow for a significant degree of control over the preparation of an experimental quantum system, provides a promising new avenue for the exploration of highly correlated systems~\cite{feynman_simulating_1982, ma_dissipatively_2019, sager_preparation_2020, sager_superconductivity_2021} and QPTs~\cite{smith_crossing_2020}.

In this work, we present a novel quantum-computing approach to quantum phase transitions that exploits the geometric structure of two-particle reduced density matrices (2-RDMs).  \tcthree{Traditional approaches to QPTs are typically framed in terms of the many-particle wave functions of ground and excited states~\cite{sachdev_quantum_2011, landau_intersection_1965, maslov_nonanalytic_2009}, which are readily prepared on hybrid quantum-classical computers but not easily measured in their entirety due to their exponential scaling with the number of particles.  Quantum computers allow us to directly probe the \tcfour{polynomially scaling} 2-RDM by tomography \cite{smart_efficient_2020} without classical storage of the \tcfour{exponentially scaling} many-particle wave function~\cite{smart_experimental_2019, sager_preparation_2020, smart_efficient_2020}, which potentially enables the treatment of significantly larger systems.  A 2-RDM-based analysis for QPTs on quantum computers could enable the study of QPTs for larger system sizes than either classical 2-RDM methods~\cite{erdahl_lower_2000, gidofalvi_computation_2006, schwerdtfeger_convex-set_2009, zauner_symmetry_2016} or wave-function-based modeling on quantum devices.}

Traditional wave function analysis of QPTs relies on finding discontinuities in the ground-state energy, but this method can miss important system symmetries. A \tc{complementary} analysis based on 2-RDMs developed by Erdahl and Jin~\cite{erdahl_lower_2000} and Gidofalvi and Mazziotti~\cite{gidofalvi_computation_2006, schwerdtfeger_convex-set_2009} characterizes QPTs in terms of the geometric set of 2-RDMs, particularly the movement of the ground-state 2-RDM between different phases. \tctwo{Separate work, inspired by the link between bipartite entanglement and QPTs, found that changes in individual elements of the 2-RDM set can also indicate critical phenomenon \cite{wu_quantum_2004}.} Additional work in this field by Zauner and Verstraete~\cite{zauner_symmetry_2016} found that a geometric analysis of the ground-state set of 2-RDMs provides a powerful visualization of symmetry breaking and phase transitions in both classical and quantum systems\tc{, which hearkens back to the geometric approach developed by Gibbs and Maxwell by generalizing Maxwell's eponymous surface to spin systems.}

{In contrast to the traditional description of phase transitions, 2-RDM theory provides a generalizable geometric framework for quantum phase transitions in terms of the convex set of 2-RDMs that has two important advantages: (1) based on a quantum information perspective, the 2-RDM theory relies upon the state space of all two-body observables rather than a specific Hamiltonian to examine the transition, and (2) it reduces the analysis of an infinite space of Hamiltonians to the study of recognizable features like planes or ruled surfaces in the finite and convex set of 2-RDMs.  Such a three-dimensional analysis allows for visualizing a greater swath of the space of all possible Hamiltonians than traditional single-order-parameter or energy-level analysis.  While sharing many of the same observables as the conventional analysis, the geometric perspective in a reduced state space creates a powerful, general framework for studying and understanding quantum criticality.}

Using the Lipkin-Meshkov-Glick (LMG) spin model~\cite{lipkin_validity_1965}, we show that even with Noisy Intermediate-Scale Quantum (NISQ) devices~\cite{wilen_correlated_2021}, the signature features of the quantum phase transition are captured from the measured set of 2-RDMs.  LMG is a widely used bench marking system for many-body approximation methods \cite{hammond_variational_2005,wahlen-strothman_merging_2017,robbins_benchmarking_2021} and has been extensively studied for its phase behavior \cite{vidal_entanglement_2004,orus_equivalence_2008,castanos_classical_2006,gidofalvi_computation_2006,ribeiro_thermodynamical_2007,heiss_largenbehaviour_2005}. Using the unitary transformations available to a quantum computer, a simulated finite-particle LMG system is manipulated through several critical regions. Tomography of the system is used to determine the ground-state set of 2-RDMs \cite{sager_preparation_2020,smart_quantum-classical_2019,smart_experimental_2019,smart_efficient_2020}, which reveals discontinuities in the system's order parameters. Additionally, the geometry of the ground-state set provides evidence of symmetry breaking on the NISQ devices \cite{gidofalvi_computation_2006, coleman_kummer_2002}.

\section{Theory}
\subsection{Reduced Density Matrix}
RDM mechanics is an alternative to the wave function mechanics traditionally used in quantum molecular studies \cite{mazziotti_realization_2004, mazziotti_quantum_2006}. Because the fundamental interactions in electronic systems are pairwise, the energies and properties of such systems are computable from a knowledge of the 2-RDM. The 1- and 2-RDMs are obtained by integrating the density matrix, $\ket{\psi}\bra{\psi}$, over $N-1$ or $N-2$ particles where the elements of the 1- and 2-RDMs are given in the second-quantization formalism by
\begin{equation}
    ^1D^i_j=\bra{\psi}\hat{a}^\dagger_i\hat{a}_j\ket{\psi}
\end{equation}
\begin{equation}
    ^2D^{ij}_{kl}=\bra{\psi}\hat{a}^\dagger_i\hat{a}^\dagger_j\hat{a}_l\hat{a}_k\ket{\psi}
\end{equation}
in which $\hat{a}_i$ and $\hat{a}^\dagger_i$ are the fermionic annihilation and creation operators for the spin-orbital $i$. These pairs of operators can be expressed as strings of Pauli matrices, which are directly measurable on the quantum computer \cite{sager_preparation_2020}. The resulting 2-RDMs, according to Rosina's theorem, completely characterize the ground-state energy and order parameters of a system with only pairwise interactions \cite{rosina_reduced_1968, mazziotti_contracted_1998}, circumventing the need for a full wave-function description of the system, which could require significantly more measurements on a NISQ device \cite{tempel_quantum_2012} increasing error and computational time. Discontinuities in the individual order parameters obtained from the 2-RDM as the system's Hamiltonian is manipulated, can be used to find critical points \cite{gidofalvi_computation_2006, coleman_kummer_2002}. \\

Three-dimensional graphical analysis of the set of 2-RDMs allows for the identification and classification of the system's critical points, as well as direct observation of symmetry breaking. Additionally, \tcfour{as these graphs are 3-D slices of the total RDM set, their} construction only requires a subset of 2-RDM elements, potentially allowing for a further pruning of necessary measurements. In the thermodynamic limit, first-order QPTs appear as planes or discontinuities in the extremal or ground-state values of the 2-RDM, which can be identified with no knowledge of the Hamiltonian of the system. \tc{In systems with more than three degrees of freedom within the RDM, maximizing the size of these discontinuous regions, by changing which 3D slice is taken, will indicate the order parameter with maximum symmetry breaking, and provides a systematic way to discover symmetry breaking in a system \cite{zauner_symmetry_2016}.} Second-order QPTs manifest as regions where the extremal values of the 2-RDM set change rapidly with changes in the Hamiltonian parameters or the curvature becomes discontinuous \cite{gidofalvi_computation_2006, zauner_symmetry_2016}. To solidify these concepts, we analyze the LMG system through the lens of RDM mechanics.

\subsection{The Lipkin-Meshkov-Glick Model}

The Lipkin-Meshkov-Glick (LMG) model system consists of two energy levels separated by $\epsilon$ each containing $N$-degenerate states. There are $N$ fermions in the system, with a coupling parameter, $\lambda$, that scatters pairs of particles between the levels. A configuration of the system is characterized by two quantum numbers: $\sigma=\pm 1$ indicating the energy level and $p=\{0,...,N\}$ specifying the state within that level \cite{lipkin_validity_1965}. In the second quantization formalism, the LMG Hamiltonian is:
\begin{multline}
   \hat{H}= \frac{1}{2}\epsilon \sum_{p\sigma} \sigma \hat{a}^{\dagger}_{p\sigma} \hat{a}_{p\sigma}+\\
    \frac{1}{2}\lambda\sum_{pp'\sigma} \hat{a}^{\dagger}_{p\sigma} \hat{a}^{\dagger}_{p'\sigma} \hat{a}_{p'-\sigma} \hat{a}_{p-\sigma} .
\end{multline}

\begin{figure}
    \includegraphics[width=9cm]{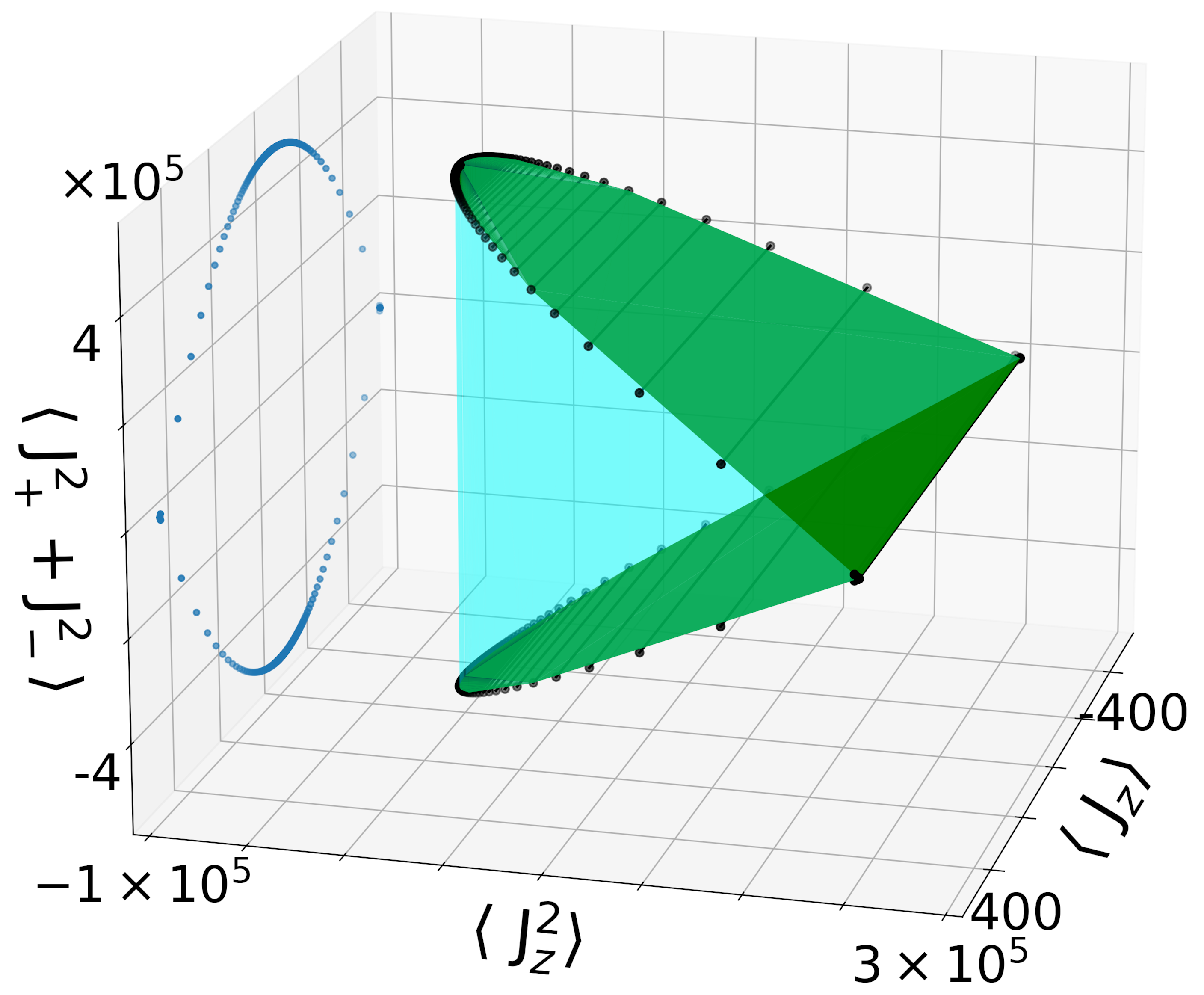}
    \caption{The Convex Hull of the 2-RDMs of the 1000-Particle Lipkin Model. The cyan and green coloring distinguish the 2 ruled surfaces of the convex set, which contain lines parallel to the $\langle \hat{J}_\text{z} \rangle$ and $\langle \hat{J}_+^2+\hat{J}_-^2 \rangle$ axes respectively. \tcthree{The line, traced by the black points, that separates the two ruled surfaces contains the set of ground-state 2-RDMs. These points are projected into the $\langle \hat{J}_\text{z}\rangle$-$\langle\hat{J}_+^2+\hat{J}_-^2 \rangle$ plane as represented by a series of blue dots.} The lines along the green surface, show the steps of a trajectory along the edges of the set as $\lambda$ is taken from infinity to zero while $\epsilon= \pm 1$. All axes are in atomic units.}
    \label{fig:3drdm}
\end{figure}

In order to solve for the LMG eigenstates exactly, it is helpful to reduce the degrees of freedom of the problem through incorporation of the system's symmetries into the Hamiltonian through a quasi-spin formalism. This formalism recognizes that the LMG two-level system is analogous \tcthree{to a system comprised of }$N$ two-spin-state particles, with a Hamiltonian:
\begin{align}
    \hat{H}=\epsilon \hat{J}_z +\frac{1}{2}\lambda(\hat{J}_+^2+\hat{J}_-^2)
    \label{ham}
\end{align}
using the traditional spin operators
\begin{align*}
   \hat{J}_+&=\sum_p \hat{a}^{\dagger}_{p,+1}\hat{a}_{p,-1},   &   \hat{J}_-&=\sum_p\hat{a}^{\dagger}_{p,-1}\hat{a}_{p,+1},
\end{align*}

and

\begin{equation}
    \hat{J}_z=\frac{1}{2}\sum_{p\sigma}\sigma \hat{a}^\dagger_{p\sigma}\hat{a}_{p\sigma}
\end{equation}

\begin{figure}
    \includegraphics[width=9cm]{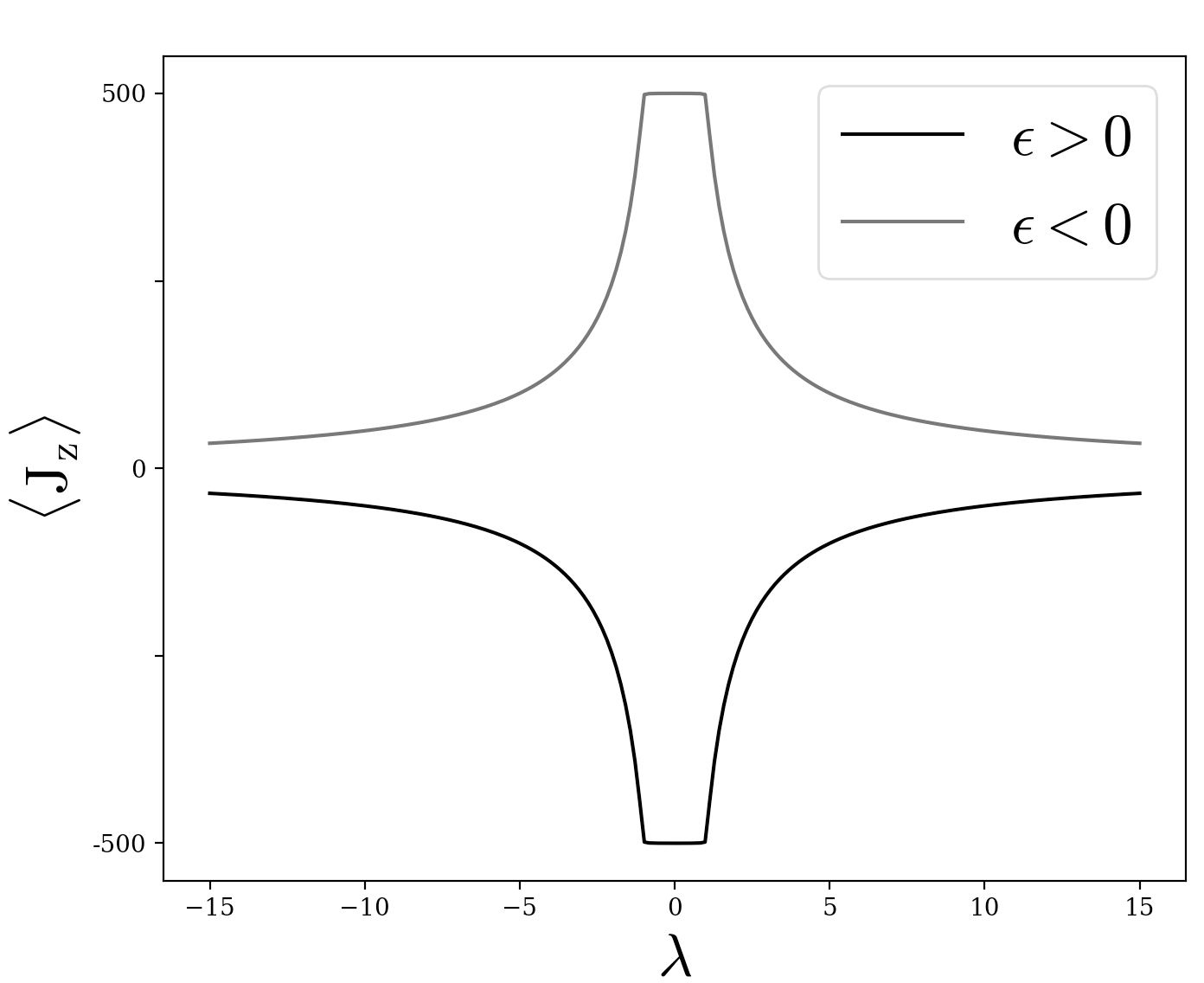}
    \caption{Symmetry Breaking of the $\langle \hat{J}_z\rangle$ Order Parameter in the 1000-Particle Lipkin Model. The black and grey lines show the change in $\langle \hat{J}_z \rangle$, in atomic units, as $\lambda$ is changed for systems with a positive and negative value of $\epsilon$, respectively.}
    \label{fig:jzsym}
\end{figure}

\begin{figure}
    \includegraphics[width=9cm]{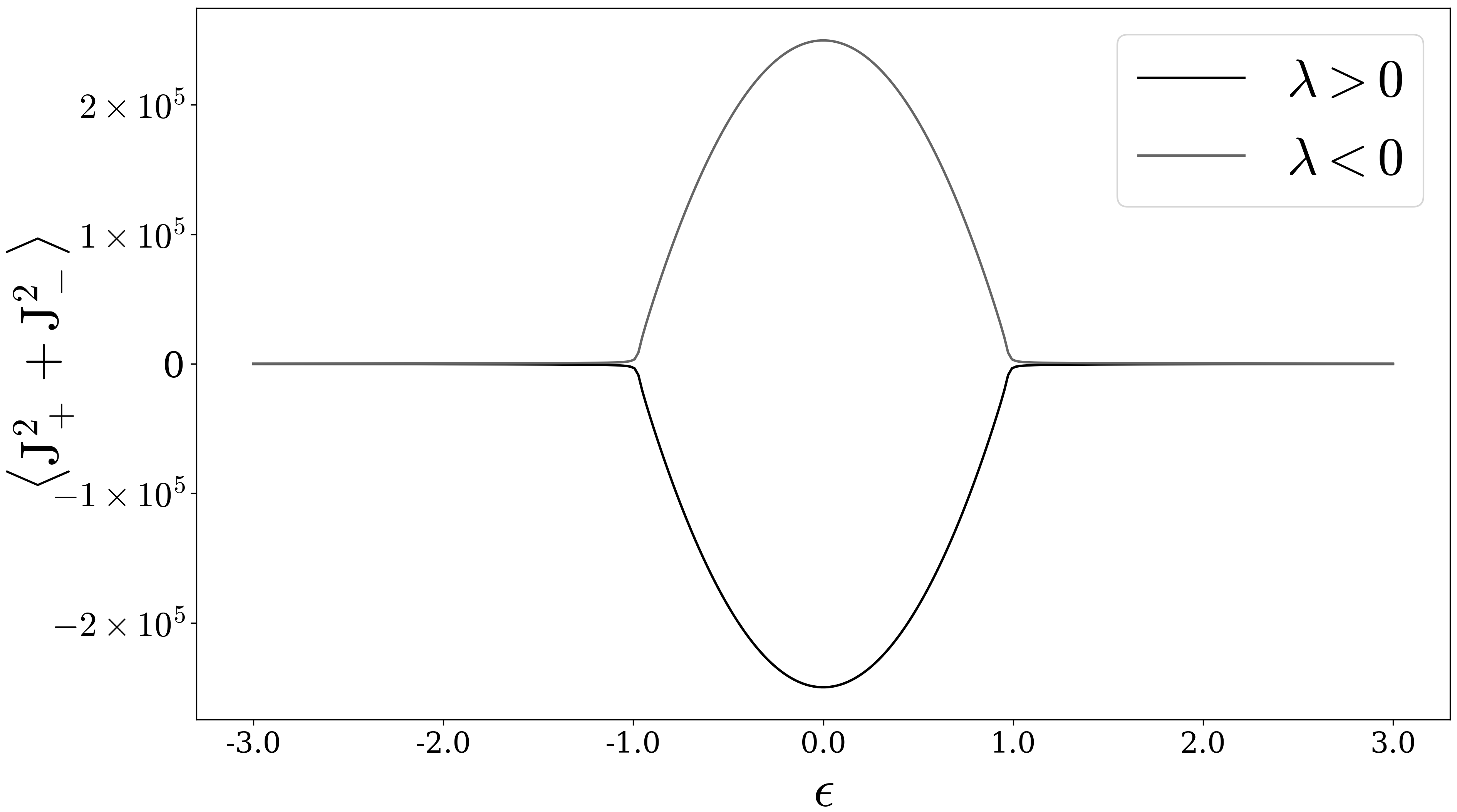}
    \caption{Symmetry Breaking of the $\langle \hat{J}_+^2+\hat{J}_-^2 \rangle$ axis in the 1000-Particle Lipkin Model. The black and grey lines show the change in $\langle \hat{J}_+^2+\hat{J}_-^2\rangle$, in atomic units, as $\epsilon$ is changed for systems with a positive and negative value of $\lambda$, respectively.}
    \label{fig:pnmsym}
\end{figure}

\begin{figure}
    \includegraphics[width=9cm]{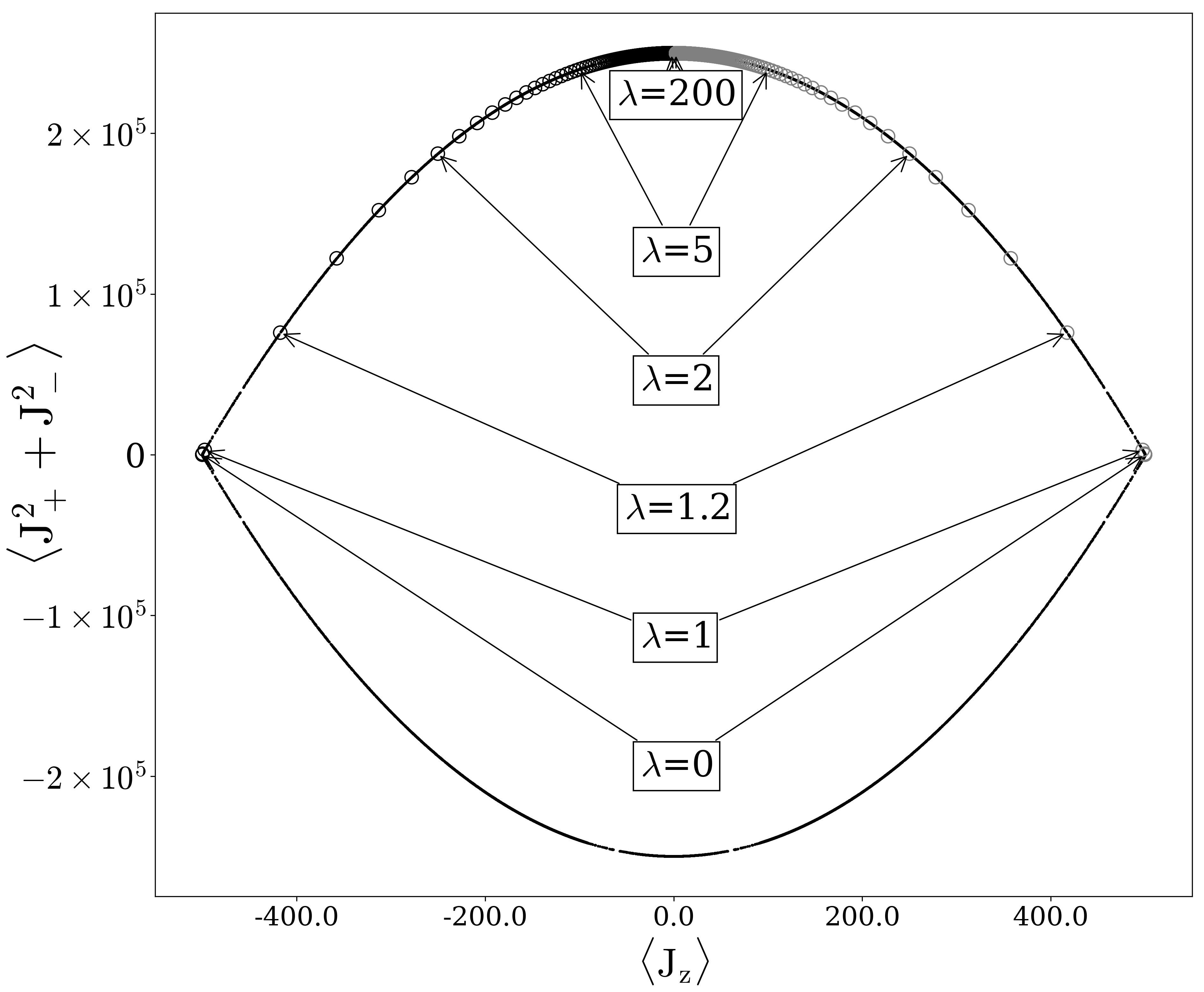}
    \caption{The 2-RDM for the 1000-Particle Lipkin Model. The black line shows the extremal or ground state values of the 2-RDM set. The black and grey circles show the trajectory of the 2-RDM along the boundary of the set as $\lambda$ is brought from infinity to zero with $\epsilon$ either being greater, for black, or less, for grey, than zero. All axes are in atomic units.}
    \label{fig:2rdm}
\end{figure}

\noindent In this quasi-spin formalism, $p$ and $\sigma$ indicate the particle and the $Z$-component of its spin, respectively. Writing the Hamiltonian in the $\ket{j,j_z}$ basis, reveals a block diagonal matrix, which can be diagonalized at a much lower computational cost than the original $2^N \times 2^N$ matrix \cite{lipkin_validity_1965}. The 2-RDM for the LMG system in this formalism is completely defined by the three expectation values $\langle \hat{J}_z\rangle$, $\langle \hat{J}_\text{z}^2\rangle$, $\langle \hat{J}_+^2+\hat{J}_-^2\rangle$ \cite{yasuda_uniqueness_2002}. Therefore the extremal points of this set can be visualized in a space defined by these parameters.

\subsection{Convex Hull of Ground-State Set of 2-RDMs}
The convex hull of the ground-state set of 2-RDMs for the 1000-particle LMG system can be seen in Fig. \ref{fig:3drdm}, which was obtained through the exact diagonalization of the Hamiltonian. This representation gives geometric insight into both general properties of the 2-RDM and more specific properties of the LMG system\tcthree{, and serves as a reference in the geometric analysis of noisier results from simulations of the LMG system on NISQ devices}. The ground-state 2-RDMs lie on the boundary of the convex set while excited-state 2-RDMs generally lie inside the convex set.  This plot also reveals the existence of two ruled surfaces, colored green and blue, indicating two forms of symmetry breaking in the system. Points along the lines of the ruled surfaces are ground states of the Hamiltonian with all of the same order parameters except for the parameter with an axis parallel to the lines. This indicates that there is symmetry breaking in the system \cite{zauner_symmetry_2016}. The green surface has lines parallel to the $\langle \hat{J}_z \rangle$ axis, which for $\epsilon \neq0$ indicates a breaking of the spin flip symmetry of the ground state in regions with small $\lambda$ \cite{stout_spontaneous_1994}. This manifests as the ground state preferring either an all up or all down spin configuration based on the value of $\epsilon$. Figure \ref{fig:jzsym} captures this symmetry breaking by demonstrating how the order parameter $\langle \hat{J}_z \rangle$ changes with $\lambda$ for systems with positive and negative values of $\epsilon$. These systems are identical for large $\lambda$ values, but as $\epsilon$ becomes a more significant contribution to the Hamiltonian, a symmetry-breaking divergence occurs between the two systems. The blue surface with lines parallel to the $\langle \hat{J}_+^2+\hat{J}_-^2 \rangle$ indicates another symmetry breaking, where the ground state has either a positive or negative eigenvalue for $\langle \hat{J}_+^2+\hat{J}_-^2 \rangle $. This eigenvalue reflects the parity of the ground-state wave function. Figure \ref{fig:pnmsym} shows the divergence of the positive and negative $\lambda$ symmetry invariant systems as the Hamiltonian is tuned from an $\epsilon$- to a $\lambda$-dominated region at $\epsilon \approx 1$. Figures \ref{fig:jzsym} and \ref{fig:pnmsym} also both contain sharp changes in the curvature at $\lambda,\epsilon \approx 1$. These sharp changes or discontinuities in the derivatives of the order parameters are signs of a second-order QPT. Signatures of these QPTs can also be found in the set of 2-RDMs.

The hull plot, Fig. \ref{fig:3drdm}, contains a projection of the 2-RDM into the $\langle \hat{J}_z \rangle$ - $\langle \hat{J}_+^2+\hat{J}_-^2 \rangle$ plane, which is seen in more detail in Fig. \ref{fig:2rdm}. This 2D representation contains two trajectories of 2-RDMs as $\lambda$ is taken from $\infty \rightarrow 0$ for positive and negative values of $\epsilon$. The plot demonstrates that the systems start from the same symmetry invariant 2-RDM at $\lambda = \infty$, but diverge radically for infinitesimally small positive and negative values of $\epsilon$. This graph also shows the presence of a second-order phase transition around $\lambda =1$. From $\lambda =0 \rightarrow 1$ the 2-RDM barely changes, but suddenly after $\lambda = 1$ the 2-RDM begins to move very rapidly along the boundary of the set, before decelerating again near the apex of the curve. This ``acceleration" of the 2-RDM is characteristic of a second-order phase transition \cite{gidofalvi_computation_2006, zauner_symmetry_2016}. For finite-particle LMG systems, this acceleration is not as rapid as it is for the system in the thermodynamic limit, where the speed of the 2-RDM diverges, but a finite signature of this QPT is still present \cite{dusuel_finite-size_2004, heiss_largenbehaviour_2005, gidofalvi_computation_2006}.

The quasi-spin formalism for the LMG system can be mapped onto the two-state qubit system of a quantum computer (QC), where the traditional computational basis is $\ket{\uparrow}=\ket{0}$ and $\ket{\downarrow}=\ket{1}$ \cite{nielsen_quantum_2010}. Therefore each quasi-particle in the LMG system or equivalently each pair of $p^{\rm th}$ states is represented by a qubit on the QC. Quantum computers also offer the ability to generate arbitrary interactions between particles including the two spin flip interaction of the LMG system (Circuit Details can be found in Appendix~\ref{app:circuit}). Measurement of the 2-RDM can then be used to identify the presence of a QPT through the methods \tcthree{previously discussed}.

\section{Results}

\begin{figure}
    \includegraphics[width=9cm]{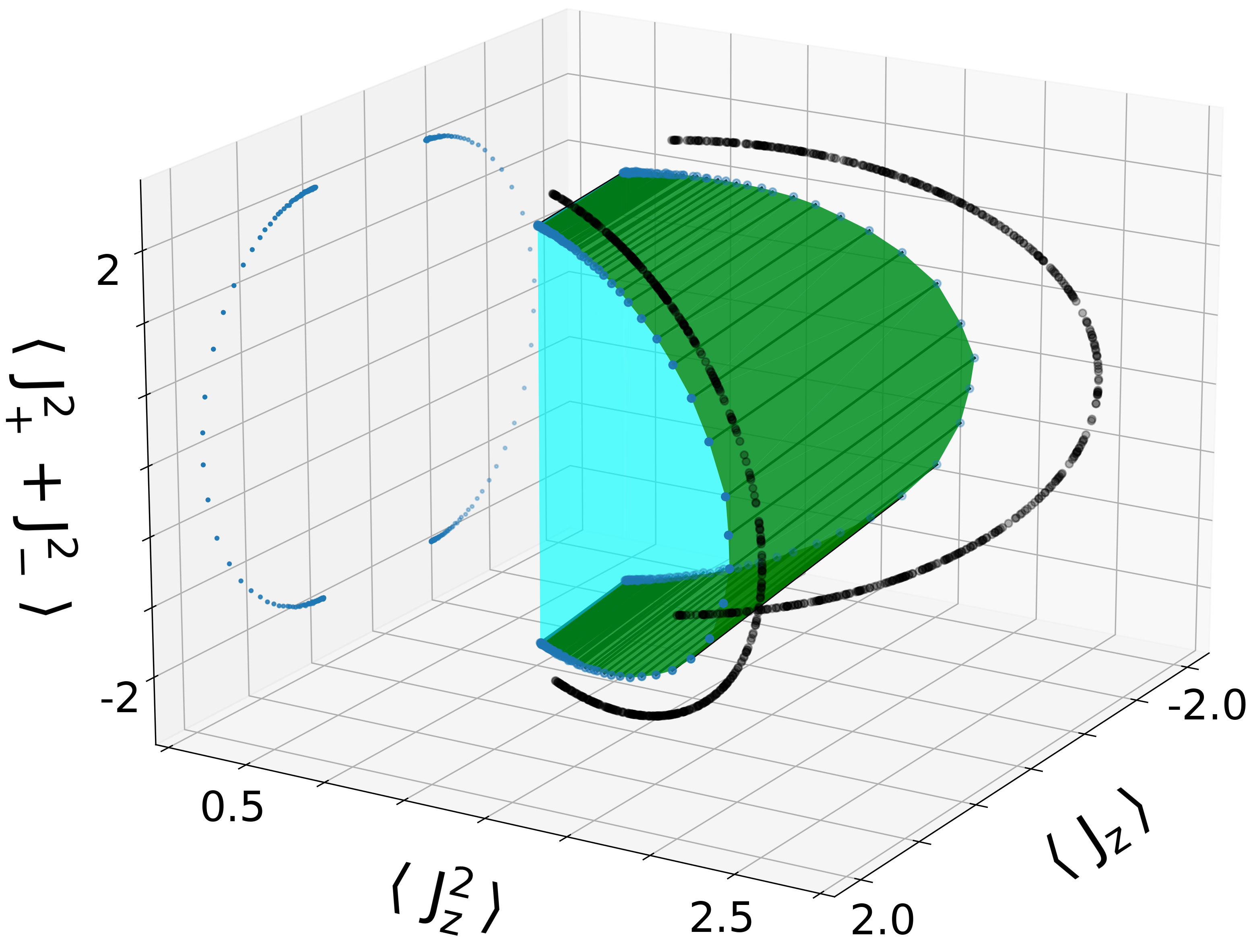}
    \caption{The Convex Hull of the 3-Particle Lipkin Model 2-RDMs. The black line outlines the edges of the exact set of 2-RDMs, while the inner shape is the convex hull of the QC results. A projection of the QC results is shown by a blue scatter plot. The cyan and green coloring distinguish the 2 ruled surfaces of the convex set. The lines along the green surface connect points of constant $\lambda$ with $\pm \epsilon$ with the distance between lines decreasing for greater values of $\lambda$. Experimental results were obtained from ibmq$\_$quito quantum computers \cite{noauthor_ibm_2021} using the circuit found in Fig. \ref{circ:3}. All axes are in atomic units.}
    \label{fig:3qu3rdm}
\end{figure}

\begin{figure}
	\includegraphics[width=9cm]{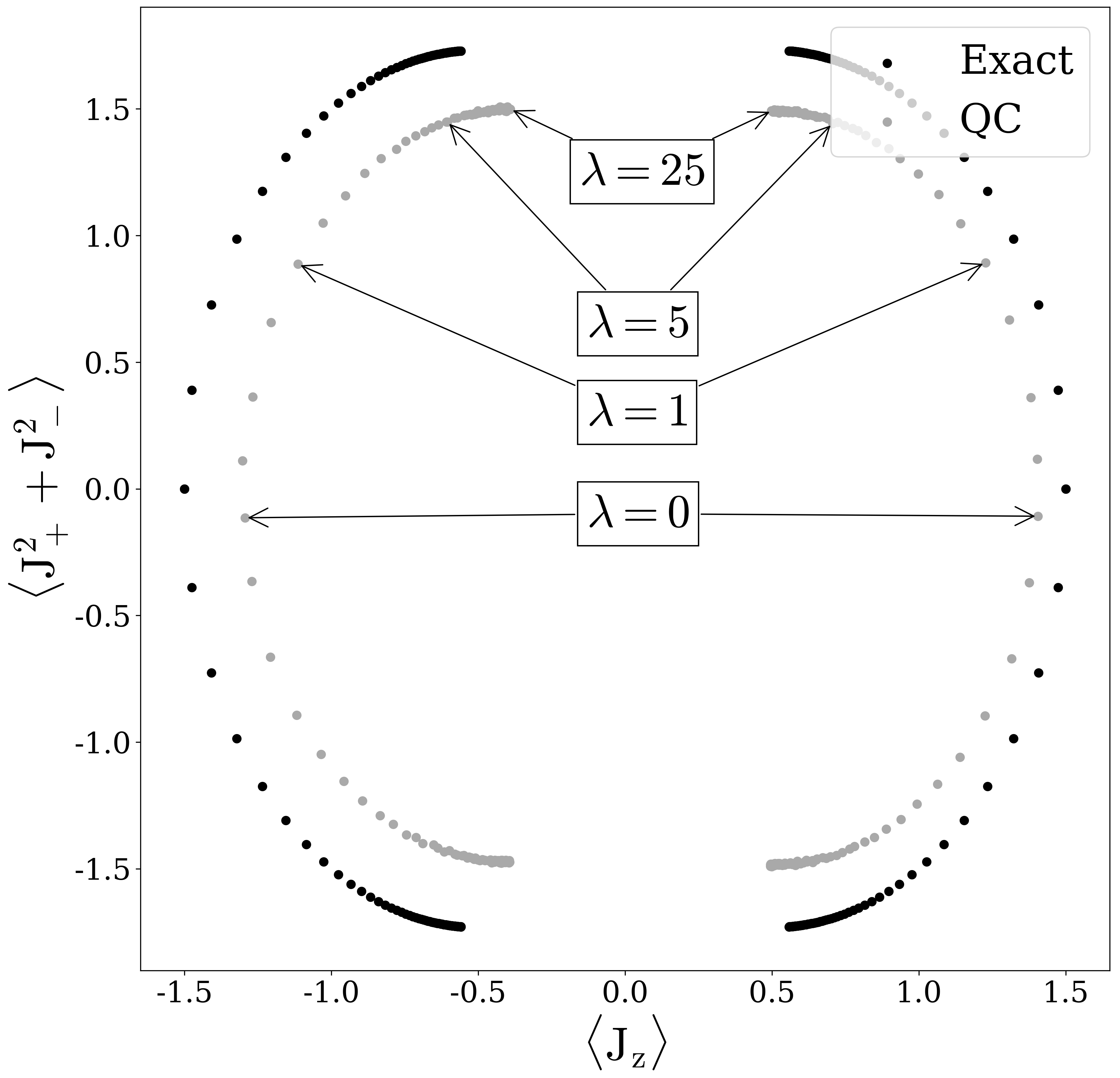}
     	\caption{Comparison of Experimental 2-RDMs with Exact for 3 qubits. Experimental results were obtained from ibmq$\_$quito quantum computer \cite{noauthor_ibm_2021}. $\lambda$, $\langle \hat{J}_z\rangle$, $\langle\hat{J}_+^2+\hat{J}_-^2\rangle$ are in reference to Eq. \ref{ham}. All axes are in atomic units.}
    	\label{fig:3pnm}
\end{figure}

\begin{figure}
	\includegraphics[width=9cm]{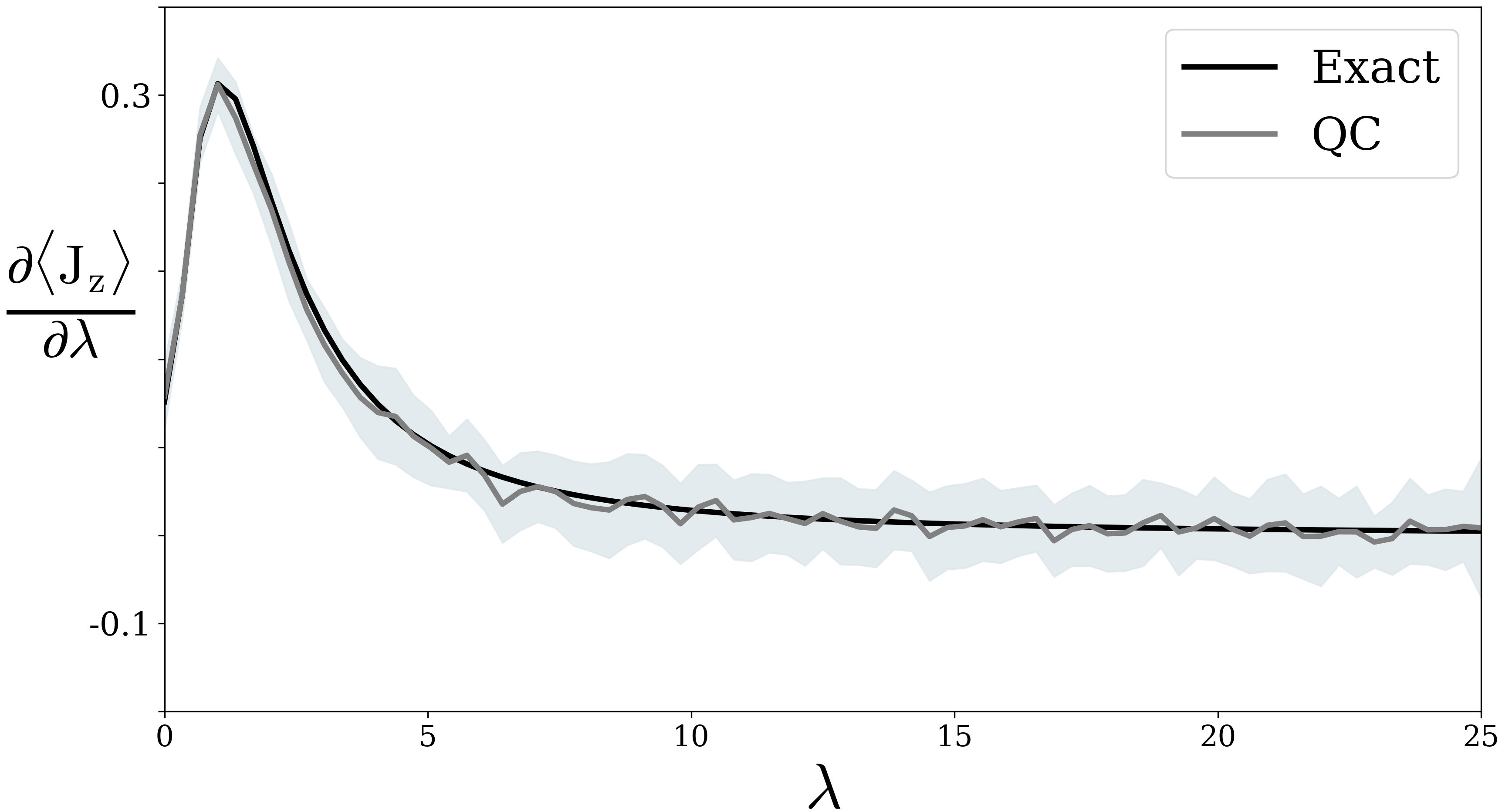}
    	\caption{Gradient of $\langle \hat{J}_z \rangle$ for Lipkin Model on 3-Qubit Circuit. \tc{Grey shading shows one standard deviation from the average of \tctwo{ten} experimental measurements \tctwo{ of 2**13 shots}. These results} were obtained from the circuit in Fig. ~\ref{circ:3} on the imbq$\_$quito quantum computer \cite{noauthor_ibm_2021}. $\lambda$ and $\langle \hat{J}_z \rangle$ are in reference to Eq. \ref{ham} and plotted in atomic units.}
    	\label{fig:3jz}
\end{figure}

Evidence of a QPT on a quantum computer is obtained from a 3-qubit simulation of the LMG system by measuring elements of the system's ground state 2-RDMs. The system is simulated by the circuit in \tcthree{Appendix}  Fig. \ref{circ:3} for $\epsilon = \pm 1$ and $\lambda\in$ [-25,25]. The experimentally gathered convex hull for this system can be seen in Fig. \ref{fig:3qu3rdm}, with the exact results outlining a larger, but similar hull. This contraction is typical in quantum computing experiments \cite{smart_experimental_2019} due to systematic errors \tctwo{like qubit crosstalk, T1 and T2 relaxation times, and gate errors. Specifically, T1 errors greatly contribute to the contraction of the set. T1 error or bit flipping populates states with lower $\langle J_z^2\rangle$ values as those states are mixed spin. $\langle\hat{J}_+^2+\hat{J}_-^2\rangle= \langle J_x^2-J_y^2\rangle$ values suffer from this same error. Measurement of $\langle J_{x/y}^2\rangle$ requires a rotation of the X or the Y component of the spin to the Z-axis (see Appendix B for details on measurement of the RDM), thus T1 relaxation populating mixed spin states will also decrease $\langle J_{x/y}^2\rangle$. Simulations on ideal quantum computers, which only include random noise, indicate that as systematic errors are decreased this contraction will also subside (additional analysis of the contraction of the set on multiple NISQ devices can be found in the SI).}
The hull, Fig. \ref{fig:3qu3rdm}, shows many of the interesting features that were present in the infinite dimensional case, Fig. \ref{fig:3drdm}, including the two ruled surfaces that indicate symmetry breaking in the system without any reference to the underlying Hamiltonian. However, in this figure a new plane parallel to the $\langle \hat{J}_z \rangle$ - $\langle\hat{J}_+^2+\hat{J}_-^2\rangle$ plane appears due to the finite size and odd number of particles in the system. The 4 vertices of this plane are degenerate 2-RDMs corresponding to the limits $(\epsilon,\lambda)\rightarrow[(0^+,>0),(0^+,<0),(0^-,>0),(0^-,<0)])$ which are all within the disordered phase of the Lipkin Model. This discontinuity in the order parameters between these degenerate states tells of a actual level and an avoided-level crossing as $\epsilon$ and $\lambda$, respectively, flip signs.

The plot contains lines of constant $\lambda$, which become more tightly spaced as $\lambda$ goes to infinity, which suggests the presence of a finite signature of a second-order QPT. This signature can be seen in Figs. \ref{fig:3pnm} and \ref{fig:3jz}. Figure \ref{fig:3pnm} demonstrates acceleration of the ground-state 2-RDMs along the boundary of the convex set of 2-RDMs between the aligned and disordered spin regions \cite{gidofalvi_computation_2006}. Despite the contraction of the set in this Figure, the change in $\langle J_z \rangle$ with respect to $\lambda$ is nearly identical to the exact results as seen in Fig. \ref{fig:3jz} showing the finite-particle signature of a second-order QPT. Note that due to the smaller particle number, the sudden acceleration at small values of $\lambda$ is less pronounced than in the thermodynamic limit shown in Fig. \ref{fig:2rdm}, but the deceleration for large values of $\lambda$ is the same.

\begin{figure}
    \includegraphics[width=9cm]{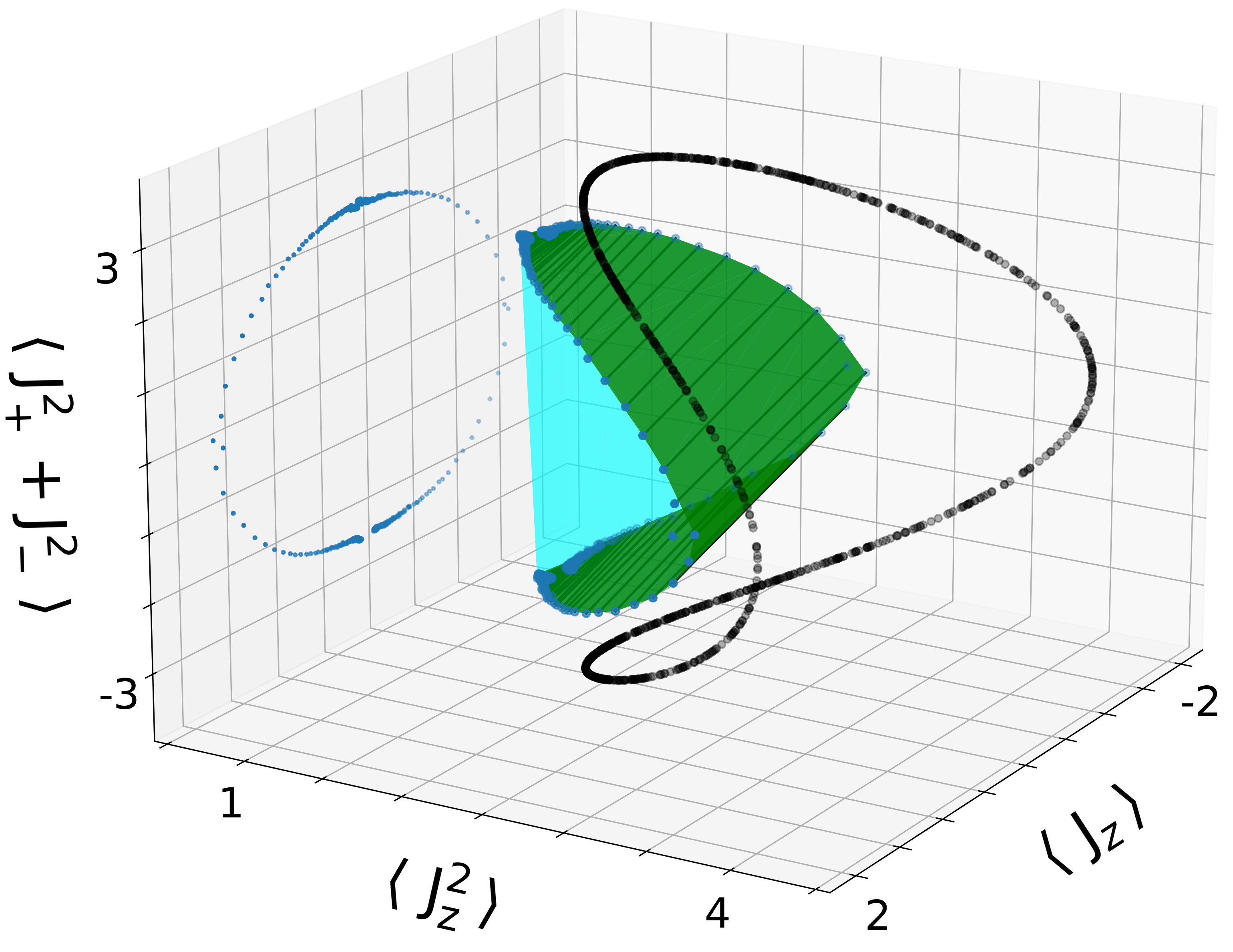}
    \caption{The Convex Hull of the 4 Particle Lipkin Model 2-RDMs. The black line outlines the edges of the exact set of 2-RDMs, while the inner shape is the convex hull of the QC results. A projection of the QC results is shown by a blue scatter plot. The cyan and green coloring distinguish the 2 ruled surfaces of the convex set. The lines along the green surface connect points of constant $\lambda$ with $\pm \epsilon$ with the distance between lines decreasing for greater values of $\lambda$. Experimental results were obtained from ibmq$\_$quito quantum computers \cite{noauthor_ibm_2021} using the circuit found in Fig. \ref{circ:4}. All axes are in atomic units.}
    \label{fig:4qu3drdm}
\end{figure}

\begin{figure}
	\includegraphics[width=9cm]{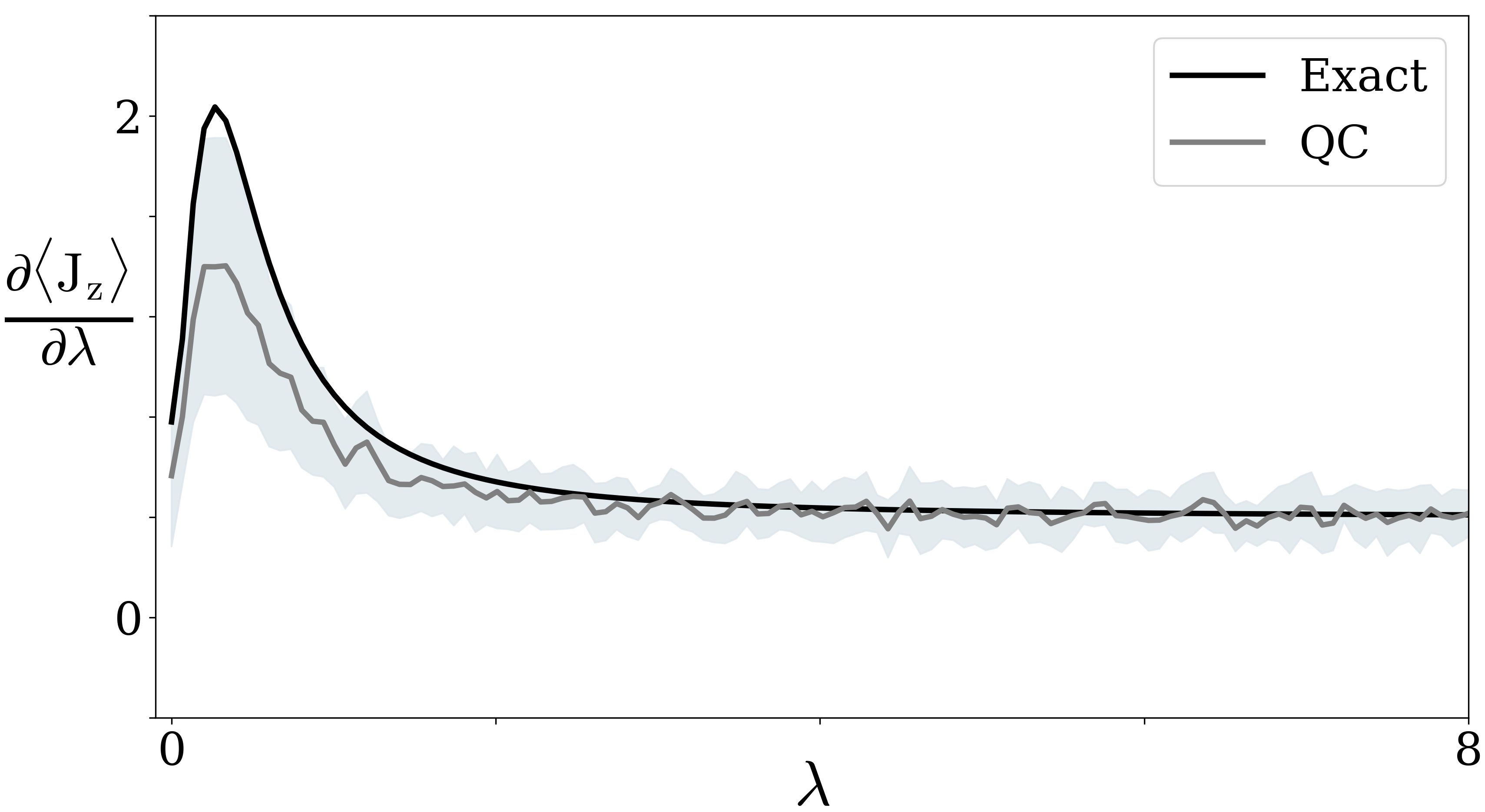}
    	\caption{Gradient of $\langle \hat{J}_z \rangle$ for Lipkin Model on 4-Qubit Circuit. \tc{Grey shading shows one standard deviation from the average of \tctwo{five }experimental measurements \tctwo{2**13 shots}. These} results were obtained from the ibmq$\_$quito quantum computer \cite{noauthor_ibm_2021} using the circuit found in Fig. \ref{circ:4}. QASM results \tc{are from }the IBM QASM simulator. $\lambda$ and $\langle \hat{J}_z\rangle$ are in reference to Eq. \ref{ham} and plotted in atomic units. }
	\label{fig:4jz}
\end{figure}

\begin{figure}
	\includegraphics[width=9cm]{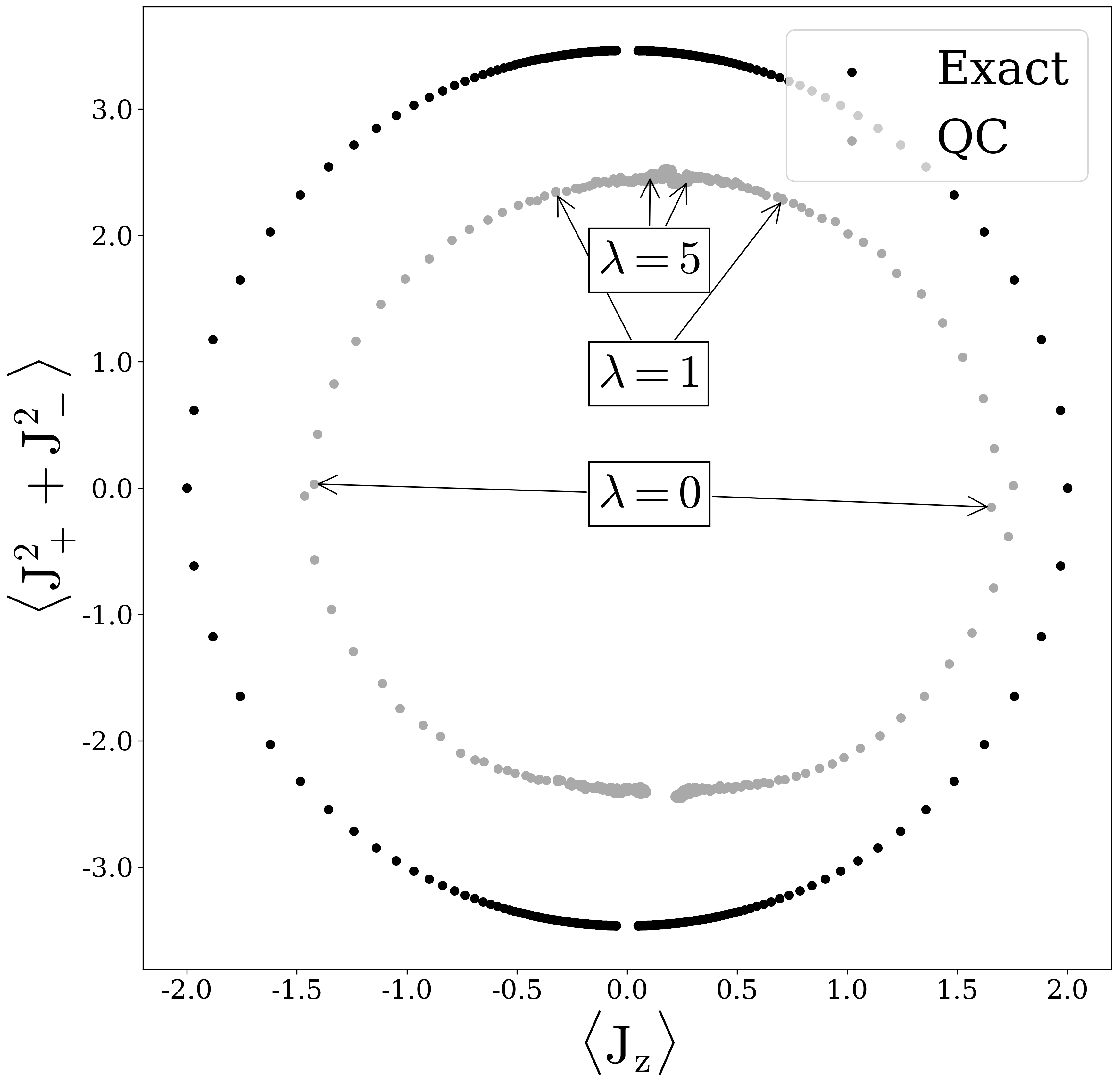}
	\caption{Comparison of Experimental 2-positive 2-RDMs with Exact for 4 qubits. Experimental results were obtained from ibmq$\_$quito quantum computer \cite{noauthor_ibm_2021}. $\lambda$, $\langle\hat{J}_z\rangle$, $\langle \hat{J}_+^2+\hat{J}_-^2\rangle$ are in reference to Eq. \ref{ham}. All axes are plotted in atomic units.}
    	\label{fig:4pnm}
\end{figure}

The convex hull of the 4-qubit 2-RDM can be seen in Fig. \ref{fig:4qu3drdm}. This hull differs immediately from the 3-qubit hull as it lacks the plane indicative of a first-order QPT due to the even number of qubits in the system. The lines along the hull illustrating the movement of 2-RDMs as $\lambda$ is increased also appear to no longer be parallel with each other or the $\langle \hat{J}_z \rangle$ axis. This asymmetry is reflected in the 2D projection which can be seen more clearly in Fig. \ref{fig:4pnm}. \tctwo{The asymmetry across the $\langle J_z\rangle$ axis arises primarily due to T1 relaxation. T1 relaxation to the ground state of the qubit, $\ket{0}$, results in a shift towards larger $\langle J_z\rangle$ values due to mapping of the quasi-spin formalism to the qubits as discussed in the theory section. T1 relaxation also explains why the effect is less drastic in the 3-qubit results, as shorter circuits are less prone to this form of decoherence. This shift has an even more dramatic effect on the $\langle J_z^2\rangle$ values as seen in Figure \ref{fig:4qu3drdm}. }The exact results in the Figure demonstrate the approach to a symmetry-invariant 2-RDM as $\lambda \rightarrow \infty$, something lacking in the 3-qubit case. The experimental results are \tctwo{ slightly} offset from one another \tctwo{but still seem to }approach a symmetry invariant 2-RDM. The characteristic acceleration and deceleration of the 2-RDMs is also apparent in the Figure. This behavior is confirmed in Figure \ref{fig:4jz}, where the data qualitatively matches the exact results peaking at $\lambda=1$, indicating the finite-particle signature of a phase transition in the system.


\section{Discussion and Conclusions}


Wave-function-based methods typically characterize QPTs by showing discontinuities in the energy surface, but can miss important symmetries of the system as well as require a \tcthree{prohibitively large} number of measurements on a noisy quantum device increasing both computational time and error. \tcthree{In contrast,} 2-RDM analysis takes advantage of the pairwise nature of interactions to decrease the number of measurements \tcthree{on a quantum computer} necessary to describe a QPT, and utilizes the geometry of the convex set to demonstrate symmetry breaking in the system without any reference to the underlying Hamiltonian. \tcthree{Here we demonstrate the 2-RDM approach to QPTs on NISQ simulators and devices by computing the finite-particle signatures of the QPT in the Lipkin model.}

\tcthree{When the scattering potential is increased in the Lipkin model, taking the system from an ordered to disordered region, the ground-state 2-RDM rapidly accelerates along the boundary of the set in the critical region. Because the 2-RDM contains information for all of the one- and two-body operators, its movement along the boundary of the set reflects the change in order parameters of the ground state, providing signatures of a QPT. Therefore, measurements showing discontinuities in this movement allow for recovery of critical behavior without any reference to the wave function~\cite{gidofalvi_computation_2006}.  The Lipkin model, shown for the 3- and 4-particle systems, has ruled surfaces on the convex hull of the set of ground-state 2-RDMs, where the lines along these ruled surfaces connect degenerate states with different values of an order parameter reflecting symmetry breaking~\cite{zauner_symmetry_2016}.}

While the Lipkin model has exact solutions, the methodology outlined could be combined with quantum eigensolvers, in particular the variational quantum eigensolver (VQE) or the contracted quantum eigensolver (CQE) for resolving ground-state 2-RDMs \cite{smart_quantum_2021}, to arrive at solutions for systems where the exact ground states are unknown from classical computations. \tcthree{NISQ devices with around 50 qubits would be able to explore systems that are well beyond the exactly solvable limit of classical devices and would strain many approximate methods. Studying real systems, like potential candidates for superconductors or exciton condensates, would be possible with VQE or CQE which are able to effectively capture long-range order due to their basis in 2-RDM theory, and being quantum chemical methods they are easily transferable to different models.} These results underscore the advantages of 2-RDM analysis on NISQ devices by demonstrating the reduced tomography costs relative to wave function methods and by utilizing the geometry of the ground-state set of 2-RDMs to resolve both symmetry breaking and phase transitions in the LMG model.

\begin{acknowledgments}
D.A.M. gratefully acknowledges the Department of Energy, Office of Basic Energy Sciences, Grant DE-SC0019215 and the U.S. National Science Foundation Grants No. DGE-1746045, No. CHE-1565638, No. CHE-2035876, and No. DMR-2037783.  We acknowledge the use of IBM Quantum services for this work. The views expressed are those of the authors, and do not reflect the official policy or position of IBM or the IBM Quantum team.
\end{acknowledgments}

\appendix
\section{Circuit Details}
\label{app:circuit}
\begin{figure}
	\centering
	\includegraphics[width=9cm]{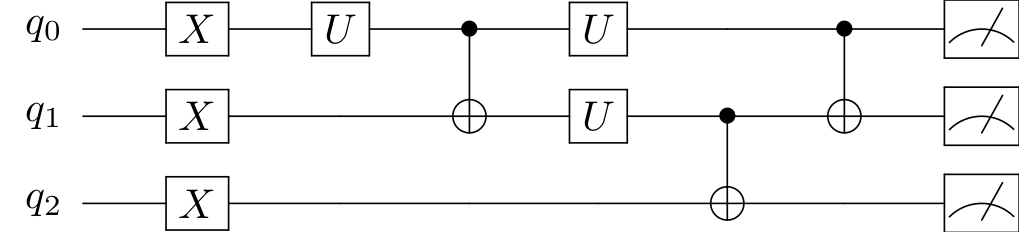}
	\caption{3-Qubit Experimental Circuit. \tc{Measurement of this circuit only provides the \emph{z}-component of the qubit expectation value. \tcthree{The determination of the \emph{x} or \emph{y} components of the expectation value requires an application of the H gate or the $S^\dagger$ then H gates, respectively, immediately before the measurement step of the circuit}. }\emph{X} is the \emph{X}-gate, \emph{U} is the traditional 2-D rotation matrix, the 2-qubit gates are CNOT gates, and the final gate is a measurement.}
	\label{circ:3}
\end{figure}

The circuits developed in this article are for the 3- and 4-particle LMG systems where the $\sigma=\pm 1$ LMG pair of spin orbitals, $p$ (see Eq \ref{ham}), are represented by the $\ket{0}$ and $\ket{1}$ states of the $p^\text{th}$ qubit respectively. \tc{The experimental expectation values result from averaging $5\times 2^{14}$ and $5\times 2^{13}$ measurements of each relevant Pauli strings (see Appendix B) on the quantum devices \tctwo{for the 3 and 4-qubit circuits respectively}. }The circuit for the 3-qubit system can be found in Fig. \ref{circ:3}. The first step is to rotate the state vector from the $\ket{000}$ configuration to the $\ket{111}$ or all down ``spin" configuration using \emph{X}-gates (following the computer science tradition in naming $\ket{0}$ the excited state) \cite{nielsen_quantum_2010}. The following unitary and CNOT gates rotate the $\ket{111}$ state into
\begin{align}
    \alpha_1 \ket{001}+\alpha_2 \ket{010}+\alpha_3 \ket{100} +\beta \ket{111}.
\end{align}
where each coefficient is a function of the rotation angles of the unitary gates. By solving the Hamiltonian exactly, the ground-state coefficients are known, so it is possible to solve a system of equations for the rotation angles. With the correct selection of rotations for the \emph{U}-gates, the resulting state is the lowest energy eigenvector of the Lipkin Hamiltonian with coefficients dependent on the parameter $\lambda$.

\begin{figure}
	\centering
	\includegraphics[width=9cm]{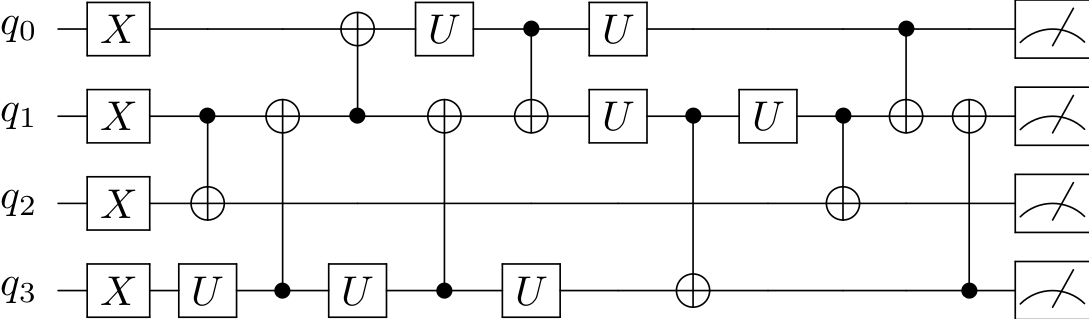}
	\caption{4-Qubit Experimental Circuit. \tc{Measurement of this circuit only provides the \emph{z}-component of the qubit expectation value. \tcthree{The determination of the \emph{x} or \emph{y} components of the expectation value requires an application of the H gate or the $S^\dagger$ then H gates, respectively, immediately before the measurement step of the circuit}.}\emph{X} is the \emph{X}-gate, \emph{U} is unitary rotation matrix (see supplemental), the 2-qubit gates are CNOT gates, and the final gates are measurements of the qubits.}
	\label{circ:4}
\end{figure}

The 4-qubit circuit, Fig. \ref{circ:4}, is constructed similarly to the 3-qubit circuit, with an initial rotation to an all ``down spin" state, and a series of rotations to the lowest energy eigenstate:
\begin{multline}
    \alpha \ket{0000}+\beta (\ket{0011}+\ket{0101}+\ket{1001}+\\
    \ket{1010}+\ket{1100})+\gamma \ket{1111}
\end{multline}

These circuits were developed with the goals of minimizing the number of CNOT gates, which are the largest sources of error in quantum computations, and satisfying the connectivity of the IBM computers ibmq$\_$belem, and imbq$\_$quito running on the Falcon r4T processor \cite{noauthor_ibm_2021}. Unlike an ideal QC, these systems can only perform CNOT gates between a subset of their qubits. CNOT gates that are not native to the machine can be decomposed into native CNOT gates, but this process can add a significant number of extra 2-qubit gates, which increases noise and hence the error of the results.

The unitary circuit referenced in Figs. \ref{circ:3} and \ref{circ:4}, can be decomposed into the basis gates:

\begin{center}
	\includegraphics[scale=.4]{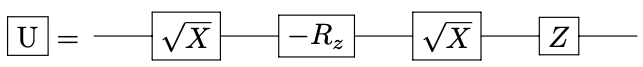}
\end{center}

\noindent In matrix notation
\begin{equation}
        \sqrt{X}=\frac{1}{2}
    \begin{bmatrix}
        1+i & 1-i \\
        1-i & 1+i
    \end{bmatrix}
    \end{equation}
    \begin{equation}
    -R_Z(\theta)=R_z(\theta+\pi)=
        \begin{bmatrix}
            1 & 0\\
            0 & e^{\theta+\pi}
        \end{bmatrix}
    \end{equation}
    \begin{equation}
        Z=R_z(3\pi)=
        \begin{bmatrix}
            1 & 0\\
            0 & -1
        \end{bmatrix}
\end{equation}
The resulting unitary matrix has the property that
\begin{align}
    U^\dagger M U=R^\dagger M R
\end{align}
where
\begin{equation}
    R=
    \begin{bmatrix}
        \cos{\theta/2} & -\sin{\theta/2}\\
        \sin{\theta/2} & \cos{\theta/2}
    \end{bmatrix}
\end{equation}
which is the classic two-dimensional rotation matrix and $M$ is any arbitrary matrix.

\section{RDM Reconstruction}
The energy expectation value of the Lipkin system can be written as a function of the 1- and 2-RDMs as
\begin{equation}
    \braket{\hat{H}}=\frac{1}{2}\epsilon \sum_{p\sigma}\sigma\text{ }^1D_p^p+\frac{1}{2} \lambda \sum_{pq}\text{ }^2D_{pq}^{pq}
\end{equation}
These elements can be constructed from linear combinations of expectation values of at most pairs of Pauli strings \cite{sager_preparation_2020} as seen below:
\begin{equation}
	^1D_p^p= \langle \sigma_{zp} \rangle
\end{equation}
\begin{equation}
	^2D_{pq}^{pq}=\langle \sigma_{xp}\sigma_{xq} \rangle -\langle \sigma_{yp}\sigma_{yq} \rangle
\end{equation}
\tc{
Figures \ref{circ:3} and \ref{circ:4} illustrate how to determine the expectation values for \emph{z}-Pauli strings, but the reconstruction of the RDM requires measurement of any combination of Pauli strings. Measuring the x and y component of a specific qubit can be accomplished by inserting a Hadamard or the adjoint phase gate ($S^\dagger$) then a Hadamard gate, respectively, before collapsing the wave function through measurement. This allows for the measurement of any combination of pairs of expectation values.
}


\bibliography{lipkin}
\end{document}